\documentclass[12pt]{article}
\usepackage{times}
\usepackage{geometry}
\usepackage{natbib}
\usepackage[utf8]{inputenc}
\usepackage{enumitem,amssymb}
\usepackage{ragged2e}
\usepackage{graphics,graphicx}
\newlist{thematic}{itemize}{8}
\setlist[thematic]{label=$\square$}
\usepackage{pifont}
\newcommand{\cmark}{\ding{51}}%

\begin{document}
\raggedright
\huge
Astro2020 Science White Paper \linebreak

\renewcommand{\baselinestretch}{1.1}
Understanding Exoplanet Atmospheres with UV Observations II: The Far UV and Atmospheric Escape \linebreak
\renewcommand{\baselinestretch}{1.0}

\normalsize

\noindent \textbf{Thematic Areas:} \hspace*{60pt} \cmark  Planetary Systems \hspace*{10pt} $\square$ Star and Planet Formation \hspace*{20pt}\linebreak
$\square$ Formation and Evolution of Compact Objects \hspace*{31pt} $\square$ Cosmology and Fundamental Physics \linebreak
  $\square$  Stars and Stellar Evolution \hspace*{1pt} $\square$ Resolved Stellar Populations and their Environments \hspace*{40pt} \linebreak
  $\square$    Galaxy Evolution   \hspace*{45pt} $\square$             Multi-Messenger Astronomy and Astrophysics \hspace*{65pt} \linebreak
  
\textbf{Principal Author:}

Name:	Eric D. Lopez
 \linebreak						
Institution:  NASA Goddard Space Flight Center
 \linebreak
Email: eric.d.lopez@nasa.gov
 \linebreak
Phone: 301.614.6951
 \linebreak
 
\textbf{Co-authors:} \linebreak
Vladimir Airapetian (NASA GSFC), Jessie Christiansen (Caltech), Luca Fossati (OeAW), Kevin France (CU Boulder)
\linebreak   

\textbf{Co-signers:} \linebreak
Daniel Angerhausen (Bern), David Ardila (JPL), Giada Arney (NASA GSFC), Vincent Bourrier (Gen\`eve), Chuanfei Dong (Princeton), Ewan Douglas (MIT/Arizona), Diana Dragomir (MIT/UNM), David Ehrenreich (Gen\`eve), Jonathan Fortney (UCSC), Adam Frank (Rochester), Mario Gennaro (STScI), Laura Kreidberg (Harvard CfA), Alain Lecavelier (IAP), Yuni Lee (NASA GSFC), Tom Louden (Warwick), Roxana Lupu (BAER/NASA Ames), Victoria Meadows (UW), Karan Molaverdikhani (MPIA), Ruth Murray-Clay (UCSC), Elisabeth Newton (Dartmouth), Erik Petigura (UCLA), Daria Pidhorodetska (NASA GSFC), Seth Redfield (Wesleyan), Aki Roberge (NASA GSFC), Leslie Rogers (Chicago), Joshua Schlieder (NASA GSFC), Adam C. Schneider (ASU), Evgenya Shkolnik (ASU), Mark Swain (JPL), Allison Youngblood (NASA GSFC)
\linebreak   

\pagebreak
\textbf{Abstract:}
Much of the focus of exoplanet atmosphere analysis in the coming decade will be at infrared wavelengths, with the planned launches of the James Webb Space Telescope (JWST) and the Wide-Field Infrared Survey Telescope (WFIRST). However, without being placed in the context of broader wavelength coverage, especially in the optical and ultraviolet, infrared observations produce an incomplete picture of exoplanet atmospheres. Scattering information encoded in blue optical and near-UV (NUV) observations can help determine whether muted spectral features observed in the infrared are due to a hazy/cloudy atmosphere, or a clear atmosphere with a higher mean molecular weight. Farther into the UV, observations can identify atmospheric escape and mass loss from exoplanet atmospheres, providing a greater understanding of the atmospheric evolution of exoplanets, along with composition information from above the cloud deck. In this white paper we focus on the science case for exoplanet observations in the far-UV (FUV); an accompanying white paper led by Jessie Christiansen will focus on the science case in the near-UV (NUV) and blue/optical.

%key points: We know atmospheric escape happens. Understanding it is important. We need UV measurements. Need big and space.

\pagebreak
%Insert your white paper text here (max of five pages including figures).

%Three figures Ehrenreich detection for GJ436b, simulated CII detection from Kevin and TESS yield summary figure.

%NOTE: All section headings are temporary, suggestions welcome.

%{\bf \noindent Introduction}

%NOTE: All section headings are temporary, suggestions welcome.

\setlength{\parindent}{2em}
\setlength{\parskip}{0.em}
%\setlength{\columnsep}{0.2em}
%\titlespacing{\section}{0pt}{0pt}{0pt}}

\section{Introduction}
%Introduce Transiting planets and environments
Transiting planets exist in extreme environments. The transit method is now the dominant mechanism of exoplanet discovery, however, due to the inherit biases of the method, most transiting planets are on short-period orbits where they are bombarded by stellar winds, flares, CMEs, and intense levels of ionizing x-ray and extreme UV radiation \citep[e.g.][]{Ribas2005,Sanz-Forcada2011,France2013,Shkolnik2014,Loyd2018}. Under these harsh conditions, planets with gas-rich atmospheres are highly vulnerable to extreme atmospheric escape \citep[e.g.][]{Lopez2012,Owen2013}. This is especially true for the abundant population of hot Neptune and sub-Neptune-sized planets identified by NASA's {\it Kepler} mission. Indeed there is strong evidence from the distribution of planets found by the {\it Kepler} mission that the transiting population out to $\sim$30 days has been strongly sculpted by escape processes, in some cases likely completely stripping the atmospheres of hot Neptunes and turning them into bare rocky super-Earths \citep[e.g.,][]{Lopez2017,Fulton2017,Owen2017}. 

\section{Why Atmospheric Escape is Important}

From the distribution of planets found by the {\it Kepler} mission, there is strong evidence that these atmospheric escape processes have played a major role in sculpting the population of short-period exoplanets \citep[e.g.][]{Baraffe2006,Lecavelier2007,Lopez2012,Owen2013,Jin2014}. There is a near complete absence of Neptune or sub-Neptunes sized planets on orbits $\lesssim$3 days \citep[e.g.][]{Howard2012,Sanchis-Ojeda2014,Lundkvist2016}, likely because any planets that formed with significant gaseous envelopes on these orbits were completely stripped by escape \citep[e.g.][]{Lopez2017}. Moreover, follow-up studies recently confirmed a clear gap in the radius distribution of {\it Kepler} planets from $\sim$1.6-2 $\mathrm{R_{\oplus}}$ \citep{Fulton2017,VanEylen2018}, a feature which was predicted by many different escape models \citep[e.g.][]{Lopez2012,Owen2013,Jin2014,Owen2017,Jin2018,Ginzburg2018}. Consequently, the compositions of the transiting planet population have been heavily impacted by escape, and any attempt to use this population to understand planet formation requires an accurate understanding of escape processes.

 Moreover, understanding atmospheric survivability is an essential component of planetary habitability. High chromospheric activity levels, such as those assumed for the young Sun, can completely strip an unshielded terrestrial atmosphere within a few Myr \citep[e.g.][]{Lammer2014,Johnstone2015}. Likewise, atmospheric escape is believed to have played a critical role in the evolution of the early Venusian atmosphere, with evidence from isotope ratios stripped up to a terrestrial ocean of water from it’s atmosphere when it was young \citep[e.g.][]{Watson1981,Kasting1983}. This is especially true for planets orbiting M dwarfs, where the habitable zone is much closer to the star and the host stars are typically much more active \citep{Audard2000,France2013,Shields2016}. Indeed some models predict that it may be extremely difficult for Earth-size planets in the habitable zone of late M dwarfs, including Proxima Cen b and the planets in the TRAPPIST-1 system, to retain an Earth-like atmosphere. \citep[e.g.][]{Luger2015,Cohen2015,Owen2016,Garcia-Sage2017,Cohen2018,Dong2018}.

\section{Background on Escape Mechanisms}

For short-period planets, there are a number of important mechanisms which can strip away significant mass from planetary atmospheres. Firstly, for planets that receive high levels of EUV and X-rays this irradiation partially ionizes hydrogen high up in a planet's atmosphere, heating gas up to $\sim10^4$ K and creating a collisional wind,  similar to outflows in HII regions, protoplantary disks, or massive stars, \citep[e.g.,][]{Parker1958, Lammer2003,Yelle2004, Murray-Clay2009,Owen2012,Koskinen2013}. Eventually, this planetary wind becomes unbound from the planet, either falling onto the star or be blown back by radiation pressure and stellar wind, in some cases forming large comet-like tails. \citep[e.g.][]{Bisikalo2013,Bourrier2016,McCann2018,Debrecht2019}.

Additionally, planets under these intense conditions are also vulnerable to escape driven by extreme space weather. Close-in planets are much more likely to be hit by large Coronal Mass Ejections, which can drive vigorous escape through ram-pressure stripping and ion exchange \citep[e.g.][]{Cherenkov2017,Khodachenko2007,Lammer2007}. Likewise, if these planets possess magnetic fields then energetic stellar wind particles may be funneled down field lines leading to significant heating of the upper atmosphere and driving escape through both thermal and non-thermal processes \citep[e.g.][]{Cohen2014,Cohen2015,Airapetian2017}.

However, despite this evidence of important role of atmospheric escape and large efforts from the theory community to build ever more complete and sophisticated models, we still lack a clear understanding of escape. Which energy sources drive escape, how do these different escape mechanisms interact, and how do these processes vary across the wide range of properties covered by both exoplanets and their host stars? To answer these questions and validate our models we need detailed observations of planetary upper atmospheres and escape.

\section{State of Current Observations}
%Introduce history of past and current observations
In principle, if we could obtain transit spectra at sufficiently high S/N and spectral resolution for multiple species we could constrain not only the overall escape rates, but also the composition, temperature, density, and velocity structure of the planetary upper atmosphere and wind \citep[e.g.][]{Bourrier2013}. Such detailed observations would be invaluable in distinguishing physics of extreme atmospheric escape, the roles of different escape mechanisms, and the impact that escape has on the exoplanet population writ large.

Indeed, this has been a major focus of HST observations over the last decade with transiting upper atmospheres now detected for a handful of exoplanets. Following the seminal detection of an exoplanet atmosphere by \citet{Charbonneau2002}, \citet{Vidal-Madjar2003} obtained STIS FUV Ly\,$\alpha$ transmission spectra of the close-in giant planet (hot Jupiter) HD209458b revealing that the planet does in fact possess a hydrodynamically escaping extended hydrogen atmosphere \citep{Lammer2003,Koskinen2010}. 

Subsequent HST FUV spectra have also revealed the presence of metals (including C, O, and Si) in this escaping atmosphere \citep{Vidal-Madjar2004, Linsky2010}, along with hydrogen escape from the hot Jupiter HD 189733b \citep{Lecavelier2012}. Meanwhile, in the NUV observations by \citet{Fossati2010} and \citet{Haswell2012} found heavy metals (Mg, Fe) escaping from the atmosphere of the extremely irradiated hot Jupiter WASP-12b. Additionally, for a handful of cases these UV observations have been supplemented with detections at other wavelengths. Using Chandra ACIS-S in soft x-rays \citet{Poppenhaeger2013} confirmed the existence of an extended partially ionized exosphere for HD 189733b. Likewise, ground based optical observations in H $\alpha$, H $\beta$ and the $\sim$517m, Mg I triplet have confirmed an extended upper atmosphere for the ultra-hot Jupiter KELT-9 b \citep{Yan2018,Cauley2019}, while \citep{Spake2018} provided the first NIR detection of atmospheric escape using Hubble's WFC3 to obtain an unresolved but high S/N transit spectrum of the 1.083 $\mu$m meta-stable Helium line in WASP-107b.

%\textbf{Luca: this is my suggestion for the previous paragraph, which I would then attach to the paragraph before. Subsequent HST observations revealed also the presence of metals (C, O, Si, and Mg) in the planetary upper atmosphere \citep{Vidal-Madjar2004, Linsky2010} ADD Vidal-Madjar et al. 2013. Additional HST observations led to the detection of an extended hydrogen envelope surrounding the hot Jupiter HD189733b (Lecavelier et al. 2012) and of heavy metal (e.g., Mg, Fe) escape in the atmosphere of the extremely irradiated hot Jupiter WASP-12b (Fossati et al. 2010; Haswell et al. 2012).}

\begin{figure}
    \centering
    \includegraphics[width=1.0\columnwidth]{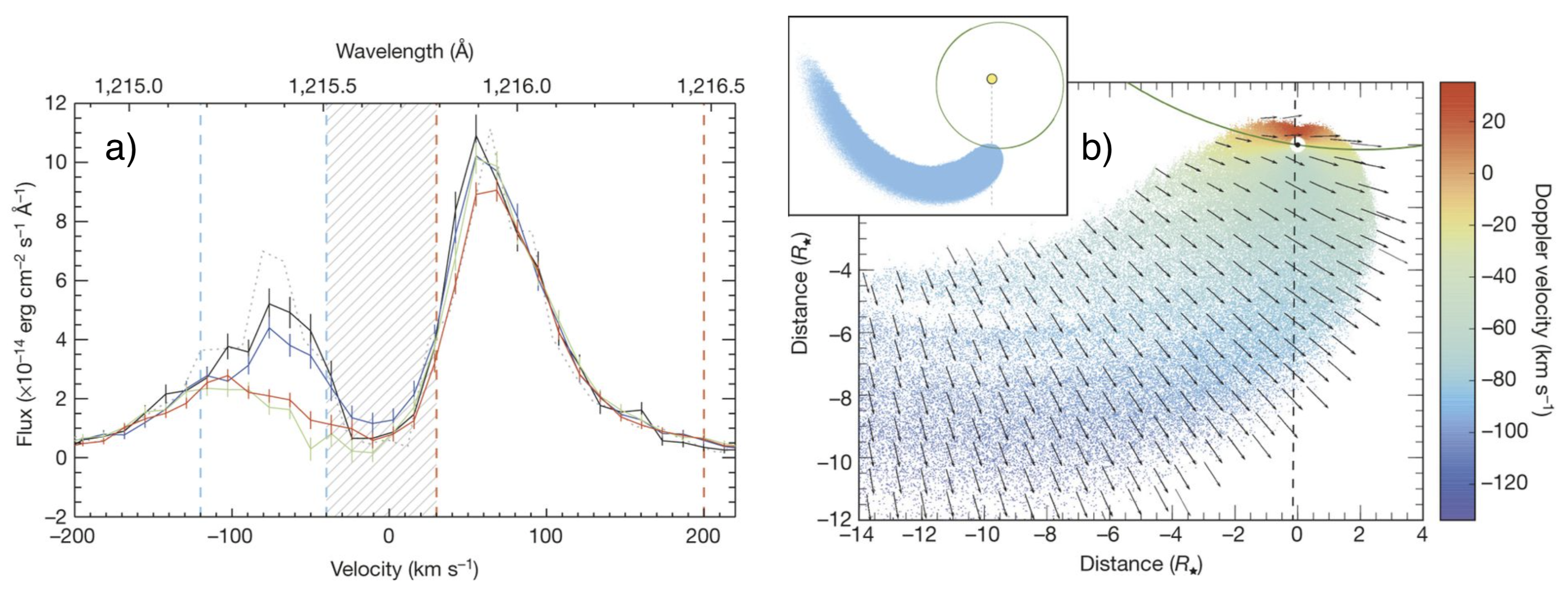}
    \caption{This shows the detection of escaping hydrogen for the nearby hot Neptune GJ 436b from \citet{Ehrenreich2015} with HST STIS. The left panel shows the Ly\,$\alpha$ velocity profile out-of-transit (black), pre-transit (blue), in-transit (green) and post-transit (red). The shading shows the low velocity region that is inaccessible due to ISM absorption. The right panel shows a reconstruction of the extended cloud of material that has escaped from the planet. This gives us an estimate of the overall escape rate, however the material that is still bound to the planet, typically at velocities $\lesssim$20 km/s is hidden by the ISM.}
    \label{fig:gj436}
\end{figure}

In recent years, these observations have been further extended down to the first Neptune-mass planets with the detection of highly extended Ly\,$\alpha$ absorption from GJ 436b \citep{Kulow2014, Ehrenreich2015} and GJ 3470b \citep{Bourrier2018}, both of which orbit nearby M dwarfs. As shown in Figure \ref{fig:gj436}, for GJ 436b these observations revealed the presence of a massive cloud of high-velocity gas from absorption in the blue-wing of the Ly\,$\alpha$ line with an enormous transit depth $\gtrsim$50\% indicating a dense comet-like tail of material that is swept up by the stellar wind and extends well beyond the planet's Hill sphere \citep{Bourrier2016}.

However, as much as we have learned from these existing observations, we are still a long way from our goal of fully characterizing planetary upper atmospheres and escape. For one thing, we still  have a very small number of confirmed escape detections and these detections have generally been limited to either hot Jupiters, or to planets orbiting M dwarfs. Even more critically, none of the existing observations have yet been able to resolve the detailed structure of the escaping upper atmosphere, particularly the critical bound portion of the thermosphere/exosphere that is of the greatest interest for constraining our models. The first major reason for this limitation is the ISM. Despite the large transit depths available in Ly\,$\alpha$, unfortunately for most stars the core of the stellar Lyman alpha line is heavily extincted by neutral hydrogen in the ISM \citep[e.g.][]{Youngblood2016}. As a result, it is typically only possible to measure Lyman alpha transits for nearby stars within $\lesssim$50pc, and even for the closest stars we can only measure absorption in the wings of the line corresponding to high velocity material that has already become unbound from the planet \citep{Ehrenreich2015}. On the other hand, while other lines at longer wavelengths have been detected in a few cases, these detections have not been at either sufficient S/N or spectral resolution to resolve the detailed structure of the thermosphere/exosphere. As a result, for all of these planets the observations only really provide a single constraint on escape processes, the overall mass-loss rate, and even that constraint is itself highly model dependent \citep[e.g.][]{Bourrier2016,Oklopcic2018,McCann2018,Shaikhislamov2018,Debrecht2019}.

\section{Why We Need the Far UV}
%Why UV is important: High spectral resolution, high temperature gas, key lines.

Access to UV wavelengths is critical to any attempt to characterize the upper atmospheres of exoplanets, star-planet interactions, or escape processes. Primarily, this is because at longer wavelengths the optical depth of the escaping material is typically too low to be detectable. Additionally, UV observations are critical for understanding the high-energy stellar heating, particularly in the extreme UV (10-90 nm), as this plays a key role in driving escape. Unfortunately, heavy ISM extinction prevents access to the $\sim$55-91 nm spectrum of nearby stars, and the 10-55 nm region requires a dedicated observatory with grazing incidence mirrors, which has not been available since EUVE \citep{Bowyer1994} ended in 2001. However, FUV spectra can be used to reconstruct the EUV stellar spectrum \citep[e.g.][]{Youngblood2016,France2018}, especially when combined with X-ray observations \citep[e.g.][]{Sanz-Forcada2011, Louden2017, King2019}. 

%While observations in some other wavelengths, such as in the 1.083um helium line and or in soft x-rays, can provide useful information on escape rates, the UV is essential both for understanding key constituents of a planet's upper atmosphere and the input stellar heating that drives escape. 

% Large HST survey programs have provided an archive of the far UV spectra of F~--~M type exoplanet host stars (e.g., France et al.  2018).  The high-energy stellar spectrum (X-ray and EUV, 5 – 900 \AA) drives and regulates atmospheric heating on short-period planets.  The high-energy stellar spectrum is a key variable in understanding the mass loss rates from these systems.   Far UV spectra can be used to estimate the EUV irradiance on these planets (Youngblood et al. 2016; France et al. 2018) or combined with X-ray observations to calculate the EUV output from a subset exoplanet host stars (Sanz-Forcada et al. 2011; Louden et al. 2017; King et al. 2018). 

Furthermore, while observations in some other wavelengths, such as in the 1.083 $\mathrm{\mu}$m helium line and or in soft x-rays can provide useful information on escape rates, the UV is essential for understanding the key constituents of a planet's upper atmosphere. The gas in the outflows of atmospheres experiencing vigorous escape is extremely hot with temperatures reaching $\gtrsim$ 10,000 K. Under these conditions any gaseous atmosphere will be broken down into a mix of atomic and ionized species, with strong lines for key species in the FUV, at $\sim118-171$ nm. In addition to Ly\,$\alpha$ at 121.6 nm, these also include major lines for CI, CII, CIV, NV, OI, SI, SiII, SiIII, and SiIV \citep{Vidal-Madjar2004,Linsky2010}. Additionally, if we wish to map the structure of the escaping wind it is important to be able to resolve the velocity profile of these lines down to the escape velocity of the planet, typically $\sim$20 km/s. This corresponds to the need for high spectral resolution with R$\gtrsim10^4$. Additionally, because FUV emission lines from host stars tend to be variable on timescales of a day or less \citep{Loyd2014}, it is important to obtain high signal to noise in any individual planet transit. This combination of requirements, high spectral resolution and high S/N in the UV, mean that these observations can only be done with large space based observatories. 
%{\bf need a paragraph here on need to characterize stellar spectrum in FUV to understand energy inputs for driving escape} \\ \\
%{\bf Stellar Influences - kf - 03/01/19  }

\section{Future Frontiers: Detecting Many Species for Many Planets}
\pagebreak

New and upcoming transit surveys like NASA's TESS \citep{Ricker2014} and ESA's PLATO \citep{Rauer2014} mission offer the promise of a large sample of transiting planets around bright nearby stars. TESS alone should find over $\sim$80 planets smaller than Neptune around stars brighter than 11th magnitude in V-band and within 50 pc \citep{Barclay2018}. These planets will be ideal for UV transmission spectra as their close proximity and bright host stars will minimize ISM extinction and maximize S/N.

With HST, it should be possible to obtain high S/N FUV transit observations to search for upper atmospheres for dozens of planets across a wide range of parameter space. Moreover, with the possibility of future large UV space observatories, we could revolutionize our understanding of upper atmospheres and escape. In addition to Ly\,$\alpha$ with larger apertures and higher throughputs, we would be able to detect transits in metal lines like OI and CII at high S/N. Due to their lower ISM extinction, these lines would allow us to observe the key low-velocity material in the lower bound portion of the upper atmosphere. Additionally, by measuring multiple independent species with different ionization temperatures, we should be able to break model-dependent degeneracies between temperature and density, allowing us to finally achieve the promise of fully mapping the structure of planetary upper atmospheres. Finally, detecting these metal species in the UV would open a unique window into understanding planet compositions in a part of the atmosphere that cannot be hidden by clouds.

\begin{figure}[t]
    \centering
    \includegraphics[width=1.0\columnwidth]{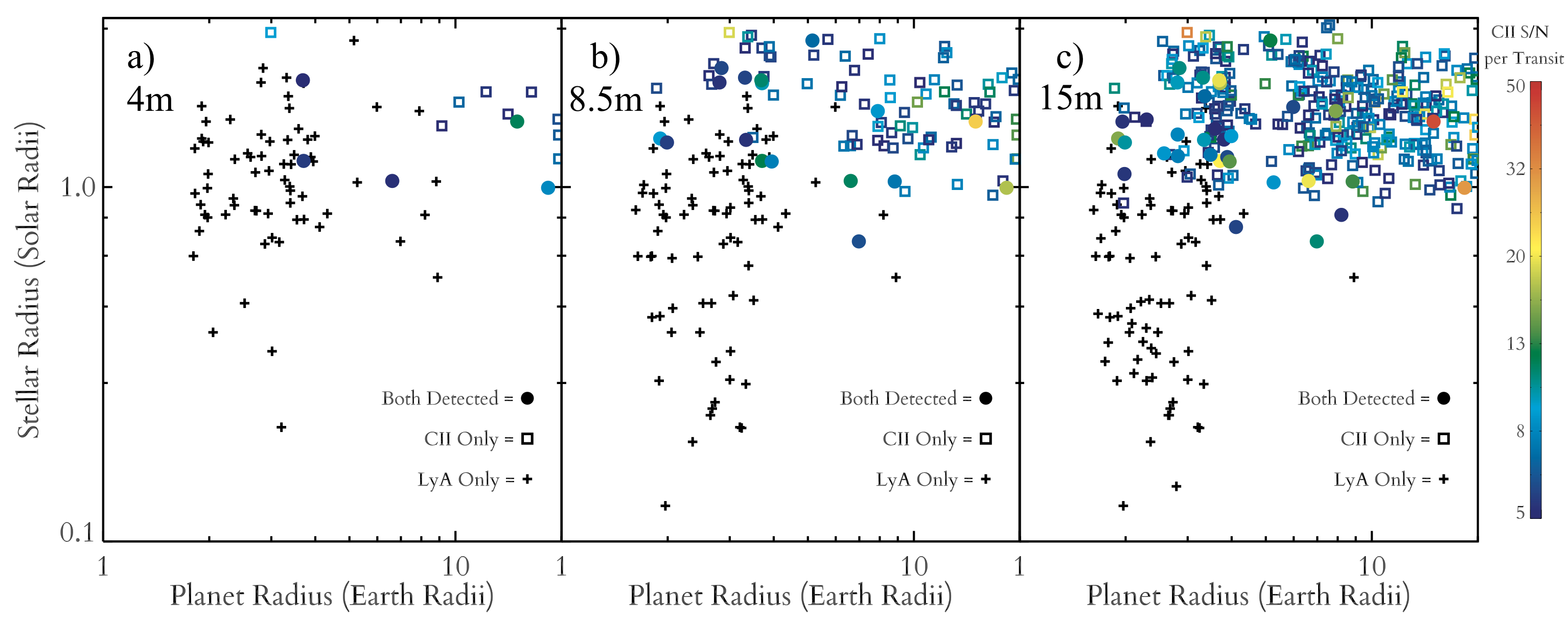}
    \caption{This shows predictions for the number of TESS planets for which we should be able to detect transiting upper atmospheres in either Ly\,$\alpha$ or the 133nm CII doublet with a UV space telescope at three different possible aperture sizes. Points show the radii of planets and their host stars for simulated TESS detections taken from \citet{Barclay2018} for which we could detect upper atmosphere absorption at $>$5$\sigma$ integrated over a single transit. The black crosses show planets where the transit can only be detected in Ly\,$\alpha$, open squares show planets where the upper atmosphere can only be detected in CII, and the filled circles show simulated planets where both  can be detected. For planets where the CII transit can be detected, points are color coded by the S/N of the planet's CII detection. All aperture sizes, including HST, should be able to detect Ly\,$\alpha$ for a wide array of planetary and stellar properties, while for apertures $\gtrsim$8 m we should also be able to fully map escaping atmospheres in FUV metal lines like OI and CII.}
    \label{fig:yield}
\end{figure}

\bigskip

\bigskip

\pagebreak
%\textbf{References}

%\bibliographystyle{apj}
%\bibliography{whitepaper}

\end{document}